\documentclass[aps,twocolumn,a4paper,10pt]{revtex4-1}
\usepackage{mathrsfs}
\usepackage{graphicx}
\usepackage{latexsym}
\usepackage{amsmath}
\usepackage{amssymb}
\usepackage{textcomp}
\usepackage{amsbsy}
\usepackage{epstopdf}

\begin{document}
 \title{\large \bf Fermion localization in a backreacted warped spacetime}
\author{Tanmoy Paul }
\email{pul.tnmy9@gmail.com}
\author{Soumitra SenGupta}
\email{tpssg@iacs.res.in}
\affiliation{Department of Theoretical Physics,\\
Indian Association for the Cultivation of Science,\\
2A $\&$ 2B Raja S.C. Mullick Road,\\
Kolkata - 700 032, India.\\}

\begin{abstract}
We consider a five dimensional AdS warped spacetime in presence of a massive scalar field in the bulk. 
The scalar field potential fulfills the requirement of modulus stabilization even when the effect 
of backreaction of the stabilizing field is taken into account. In such a scenario, we explore the role of backreaction  
on the localization of bulk fermions which in turn determines the effective radion-fermion coupling 
on the brane. Our result reveals that both the chiral modes 
of the zeroth Kaluza-Klein (KK) fermions get localized near TeV brane as the backreaction of the scalar field increases. 
We also show that the profile of massive KK fermions shifts towards the Planck brane with increasing backreaction parameter.
\end{abstract}
\maketitle

\section{Introduction}
Eversince the original proposal of Kaluza-Klein (KK) regarding the existence of extra spatial 
dimension(s) it is often believed that our  universe is  a 3-brane embedded in a higher dimensional 
spacetime and  is described  through  a low energy 
effective theory on the brane carrying the signatures of extra dimensions \cite{kanno,shiromizu}

Till date, Standard Model (SM) of particle Physics is the best suited model 
for describing the possible interactions between fundamental particles up to 
TeV scale. However, the Standard Model carries a quadratic divergence in 
the radiatively corrected  Higgs mass which can be set to the desired value $~ 126$ GeV only 
through an unnatural fine tuning of the parameters of the theory.
Among various models proposed  to solve this fine tuning problem  
\cite{arkani,horava,RS1,kaloper,cohen,burgess,chodos}
the Randall-Sundrum (RS) warped geometry model \cite{RS1} earns a special attention since 
it resolves the gauge hierarchy problem without introducing any intermediate scale between 
Planck and the  TeV scale. The interbrane separation, known 
as modulus (or radion), is assumed to be 
$\sim$ Planck length in order to generate the required hierarchy between the branes.\\
A suitable potential with a stable minimum 
is therefore needed for modulus stabilization. Goldberger 
and Wise (GW) proposed a useful stabilization mechanism \cite{GW} by introducing  
a massive scalar field in the bulk with appropriate boundary values. Not only, the 
stable value of the modulus appear as a parameter in the low 
energy effective theory on the brane, but it's fluctuation about that stable value leads 
to dynamical modulus (or radion) field which couples to the fields on the 
observable brane. This resulted into
a large volume of work on phenomenological and cosmological implications \cite{GW_radion,kribs,julien,wolfe,sumanta,sorbo} 
of modulus field  in RS warped geometry model.\\
Though the backreaction 
of the stabilizing scalar field was originally neglected  
in GW proposal, its  implications are subsequently studied in \cite{kribs, tp2}. It has been demonstrated in \cite{tp2} 
that the modulus of RS scenario can be stabilized using GW prescription even by incorporating the  
backreaction of the stabilizing field. In such a braneworld scenario, several models were 
proposed by placing the standard model fields inside the bulk .
Specially the  localization property of a bulk fermion field \cite{bajc,smyth,chang,koley,ssg1,grossman,jm,ssg}
has been a subject of great interest where explanations for observed chiral nature of massless 
fermion and the hierarchial masses of fermions among different generations have been explored. 
In this context, it is observed that the overlap of the bulk fermion wave function on  
our visible brane plays crucial role in determining the effective radion-fermion coupling 
which in turn determines phenomenology of the radion with brane matter fields.

 In view of above, it is worthwhile to address the role of backreaction of the stabilizing field 
 on fermion localization. We aim to address this in the present work.

Our paper is organized as follows : The backreacted RS scenario 
and its modulus stabilization is described in section II. Section III addresses the localization 
property of bulk fermion field and its consequences. The paper ends with some concluding remarks.

\section{Backreacted RS model and its modulus stabilization}
The action for the RS geometry with a stabilizing scalar field $\Phi$ \cite{kribs} is as follows,
\begin{eqnarray}
 S&=&\int d^5x \sqrt{G}[-M^3R+\Lambda]\nonumber\\
&+&\int d^5x \sqrt{G}[(1/2)G^{MN}\partial_M\Phi\partial_N\Phi-V(\Phi)]\nonumber\\
 &-&\int d^4x \sqrt{-g_{hid}}\lambda_{hid}(\Phi) - \int d^4x \sqrt{-g_{vis}}\lambda_{vis}(\Phi)
 \label{action}
\end{eqnarray}
where $M$ is the five dimensional Planck scale, $\Lambda$ is the bulk cosmolgical constant, 
$G_{MN}$ is the 
five dimensional metric, $g_{hid}$ and $g_{vis}$ are the induced 
metric on hidden and visible brane respectively.  
 $\lambda_{hid}$, $\lambda_{vis}$  are the self interactions of scalar field (including brane 
tensions) on Planck and TeV branes. 
The background metric ansatz is given by,
\begin{equation}
 ds^2 = \exp{[-2A(y)]}\eta_{\mu\nu}dx^{\mu}dx^{\nu} - dy^2
 \label{metric ansatz}
\end{equation}
where $A(y)$ is the warp factor. The bulk scalar field is 
assumed to depend only on the extra dimensional coordinate ($y$) \cite{kribs}.

In order to get an analytic 
solution of the backreacted geometry, the form of the scalar field potential is chosen  
as \cite{kribs},
\begin{equation}
 V(\Phi) = (1/2)\Phi^2(u^2+4uk) - (\kappa^2/6)u^2\Phi^4
 \label{scalar potential}
\end{equation}
where $k=\sqrt{-\kappa^2\Lambda/6}$. The potential contains quadratic
 as well as quartic self interaction of the scalar field and the two terms are connected 
 by a common free parameter '$u$'.
Using this form of the potential, one obtains a 
solution of coupled gravitational-scalar field equations as:
\begin{eqnarray}
 A(y) = k|y| + (\kappa^2/12)\Phi_{P}^2\exp{(-2u|y|)}
 \label{solution of warp factor}\\
 \Phi(\varphi) = \Phi_{P}\exp{(-u|y|)}
 \label{solution of scalar field}
\end{eqnarray}
where $\Phi_{P}$ is the value of the scalar field on Planck brane. 
From eqn.(\ref{solution of warp factor}), it can be argued that 
$\kappa\Phi_P$ controls the deviation of the warp factor from RS model and thus $\kappa\Phi_P$ 
is known as scalar field backreaction parameter.
 Moreover $\lambda_{hid}$ and $\lambda_{vis}$ can be obtained from the 
boundary conditions on branes as,
\begin{eqnarray}
 \lambda_{hid} = 6k/\kappa^2 - u\Phi_{P}^2
 \label{planck brane tension}\\
 \lambda_{vis} = -6k/\kappa^2 + u\Phi_{P}^2\exp{(-2u\pi r_c)}
 \label{planck brane tension}
 \end{eqnarray}
 
 where $r_c$ is the compactification radius of the extra dimension. 
 Once the solutions of $A(y)$ and $\Phi(y)$ are obtained (see eqn.(\ref{solution of warp factor}) 
 and eqn.(\ref{solution of scalar field})), the 
 modulus can be stabilized using GW prescription. It has been 
 demonstrated in \cite{tp2} that the interbrane separation in backreacted RS scenario is stabilized 
 at a value given by, 
 \begin{equation}
   k\pi r_c = \frac{k}{u}\ln{\{\frac{\kappa\Phi_{P}}{2\sqrt{1+\frac{2k}{u}}}\}}
   \label{stabilized modulus}
  \end{equation}
  
  After stabilizing the modulus of RS scenario with the modification due to backreaction 
  of the stabilizing scalar field, 
   we now show how the backreaction parameter ($\kappa\Phi_P$) affects the localization of fermion field 
  within the five dimensional spacetime.
  
  \section{Fermion localization}
  Consider a bulk massive fermion field propagating on a background geometry model 
  characterized by action in eqn. (\ref{action}). The lagrangian for the Dirac fermions is 
  given by 
  \begin{equation}
   L_{Dirac} = e^{-4A(y)} [\bar{\Psi}i\Gamma^{a}D_{a}\Psi - m_5\bar{\Psi}\Psi]
   \nonumber
  \end{equation}
  where $\Psi=\Psi(x^{\mu},y)$ is the fermion field and $m_5$ is its mass. 
  $\Gamma^{a} = (e^{A(y)}\gamma^{\mu}, -i\gamma^{5})$ denotes the five dimensional gamma 
  matrices where $\gamma^{\mu}$ and $\gamma^{5}$ represent 4D gamma matrices in chiral 
  representation. Curved gamma matrices obey the Clifford algebra i.e. $[{\Gamma^{a},\Gamma^{b}] = 2G^{ab}}$. 
  The covariant derivative $D_{a}$ can be calculated by using the metric in eqn. (\ref{metric ansatz}) 
  and is given by 
  \begin{eqnarray}
   &D_{\mu} = \partial_{\mu} - \frac{1}{2}\Gamma_{\mu}\Gamma^{4}A'(y)e^{-A(y)}\nonumber\\
   &D_{5} = \partial_{y}
   \nonumber
  \end{eqnarray}
  Using this set up, the Dirac lagrangian $L_{Dirac}$ turns out to be,
  \begin{eqnarray}
   L_{Dirac}&=&e^{-4A(y)} \bar{\Psi}[ie^{A(y)}\gamma^{\mu}\partial_{\mu}\nonumber\\
   &+&\gamma^{5}(\partial_{y}-2A'(y)) - m_5]\Psi
   \label{dirac lagrangian}
  \end{eqnarray}
  
  We decompose the five dimensional spinor via Kaluza-Klein (KK) mode expansion as 
  $\Psi(x^{\mu},y) = \sum \chi^n(x^{\mu})\xi^n(y)$, where 
  the superscript $n$ denotes the nth KK mode.
  $\chi^n(x^{\mu})$ is the projection of $\Psi(x^{\mu},y)$ on the 3-brane and $\xi^n(y)$ 
  is the extra dimensional component of 5D spinor. Left ($\chi_L$) and right ($\chi_R$) 
  states are constructed by $\chi^n_{L,R} = \frac{1}{2}(1 \mp \gamma^{5})\chi^n$. 
  Thus the KK mode expansion can be written in the following way : 
  \begin{equation}
   \Psi(x^{\mu},y) = \sum[\chi_L^{n}(x^{\mu})\xi_L^{n}(y) + \chi_R^{n}(x^{\mu})\xi_R^{n}(y)]
   \label{kk mode expansion}
  \end{equation}
  Substituting the KK mode expansion 
  of $\Psi(x^{\mu},y)$ in the Dirac field lagrangian given in eqn.(\ref{dirac lagrangian}), 
  we obtain the following equations of motion for $\xi_{L,R}(y)$ as follows :
  \begin{eqnarray}
   e^{-A(y)}&[&\pm(\partial_{y} - 2A'(y)) + m_5]\xi^n_{R,L}(y)\nonumber\\ 
   &=&-m_n\xi^n_{L,R}(y)
   \label{eom for wave function}
  \end{eqnarray}
  
  where $m_{n}$ is the mass of nth KK mode. The 4D fermions obey the canonical 
  equation of motion $i\gamma^{\mu}\partial_{\mu}\chi^n_{L,R} = m_{n}\chi^n_{L,R}$. 
  Moreover eqn.(\ref{eom for wave function}) is obtained provided the following normalization conditions 
  hold :
  \begin{equation}
   \int_{0}^{\pi} dy e^{-3A(y)}\xi_{L,R}^{m}\xi_{L,R}^{n} = \delta_{m,n}
   \label{norm1}
  \end{equation}
  \begin{equation}
   \int_{0}^{\pi} dy e^{-3A(y)}\xi_{L}^{m}\xi_{R}^{n} = 0
   \label{norm2}
  \end{equation}
  
  In the next two subsections, we discuss the localization scenario for massless 
  and massive KK modes respectively.
  
  \subsection{Massless KK mode}
  For massless mode, equation of motion of $\xi_{L,R}$ takes the following form (taking $\frac{\kappa\Phi_P}{\sqrt{2}}=l$)
  \begin{eqnarray}
   &\exp&{(-ky-\frac{l^2}{6}e^{-2uy})} [\pm(\partial_{y} - 2k + \frac{2l^2u}{3}e^{-2uy})\nonumber\\ 
   &+&m_5]\xi_{R,L}(y) = 0
   \label{eom for wave function1}
  \end{eqnarray}
  
  where we use the form of warp factor i.e. $A(y) = ky+\frac{l^2}{6}e^{-2uy}$. 
  Solution of eqn.(\ref{eom for wave function1}) is given by :
  \begin{eqnarray}
   \xi_{L,R}(y)&=&\sqrt{\frac{k}{e^{2k\pi r_c}-1}}[1 + \exp{(\frac{2l^2}{3}e^{-u\pi r_c})}]^{1/2}\nonumber\\
   &*&\exp{(\frac{l^2}{6}e^{-2uy})} e^{2ky}
   \label{solution1}
  \end{eqnarray}
  for $m_5 = 0$; and
  
  \begin{eqnarray}
   \xi_{L,R}(y)&=&\sqrt{\frac{k}{e^{(2k\pm m_5)\pi r_c}-1}}[1 + \exp{(\frac{2l^2}{3}e^{-u\pi r_c} \mp \frac{2m_5}{k})}]^{1/2}\nonumber\\
   &*&\exp{(\frac{l^2}{6}e^{-2uy})} e^{(2k \pm m_5)y}
   \label{solution2}
  \end{eqnarray}
  for $m_5 \neq 0$.
  
  The overall normalization constants in eqn. (\ref{solution1}) and eqn. (\ref{solution2}) 
  are determined by using the normalization condition presented earlier in eqn. (\ref{norm1}). 
  It may be noticed that left and right chiral modes have the same solution when $m_5=0$, 
  but the degeneracy between the two chiral modes lifted in the presence of 
  non-zero bulk fermionic mass term.
  
  It is worthwhile to study how the localization scenario depends on the 
  backreaction parameter (l) as well as the bulk mass parameter ($m_5$).
  
  \subsubsection*{Effect of backreaction parameter}
\begin{figure}[!h]
\begin{center}
 \centering
 \includegraphics[width=3.0in,height=2.0in]{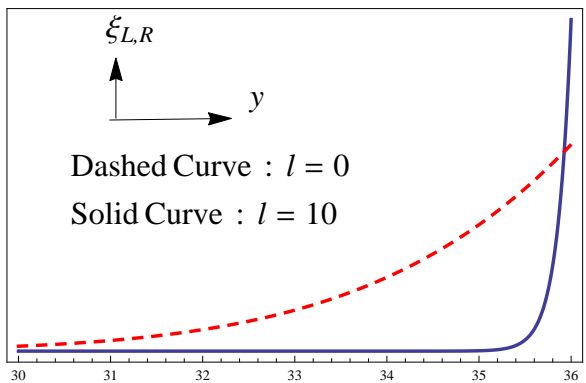}
 \caption{$\xi_{L,R}$ vs $y$ for $k=1$, $\frac{u}{k}=0.2$ and $m_5=0$}
 \label{plot zero bulk mass various l zoom}
\end{center}
\end{figure}

  From eqn. (\ref{solution1}), we obtain the Figure (\ref{plot zero bulk mass various l zoom}) between $\xi_{L,R}$ and $y$ for various 
  values of backreaction parameter.\\
The constant $y$ hypersurfaces at $y=0$ and $y=36$ represent Planck and TeV branes respectively. 
We focus into the region near the TeV brane (see figure (\ref{plot zero bulk mass various l zoom})) 
to depict the localization properties of the left and right modes.

Figure (\ref{plot zero bulk mass various l zoom}) clearly demonstrates that for $m_5=0$, the two chiral modes 
get more and more localized on TeV brane as the backreaction parameter comes close to $l=25$ (needed 
to solve the gauge hierarchy problem for $\frac{u}{k}=0.2$) from a lower value. Thus the solution of 
hierarchy problem and the localization of fermion are linked with each other. The figure also reveals that 
larger the backreaction parameter $l$, the localization of both the chiral modes of massless fermions becomes sharper near the 
visible brane.

On the other hand, for small values of $l$, the fermions are clearly 
localized deep inside the bulk spacetime and without any backreaction (i.e. $l=0$) 
we retrieve the RS solution where the fermions are 
peaked away from the visible brane. Thus without any bulk mass term, the fermions can 
be localized at different regions inside the bulk by adjusting the value of backreaction 
parameter.\\

From eqn. (\ref{solution2}), we obtain the plots of left and right chiral modes for various $l$ 
in presence of non-zero bulk fermionic mass.

\begin{figure}[!h]
\begin{center}
 \centering
 \includegraphics[width=3.0in,height=2.0in]{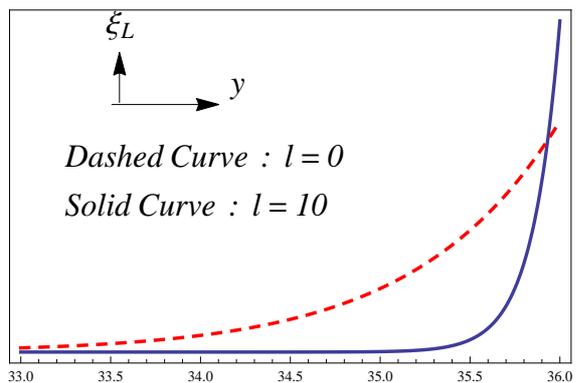}
 \caption{$\xi_{L}$ vs $y$ for $k=1$, $\frac{u}{k}=0.2$ and $m_5=0.5k$}
 \label{plot bulk mass left various l zoom}
\end{center}
\end{figure}

\begin{figure}[!h]
\begin{center}
 \centering
 \includegraphics[width=3.0in,height=2.0in]{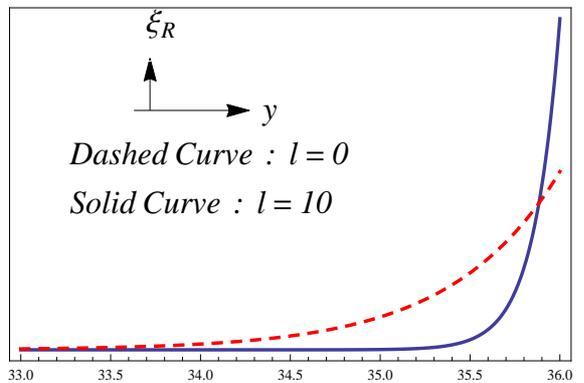}
 \caption{$\xi_{R}$ vs $y$ for $k=1$, $\frac{u}{k}=0.2$ and $m_5=0.5k$}
 \label{plot bulk mass right various l zoom}
\end{center}
\end{figure}

Figure (\ref{plot bulk mass left various l zoom}) and figure (\ref{plot bulk mass right various l zoom}) 
reveal that as the backreaction parameter increases, the peak of both left and right chiral wave 
function get shifted towards the visible brane.\\

Moreover, using the solution of $\xi_{L,R}(y)$ (in eqn. (\ref{solution2})), we obtain the effective coupling \cite{rizzo} between 
radion and zeroth order fermionic KK mode as follows:
 \begin{eqnarray}
   \lambda_L&=&\sqrt{\frac{k}{24M^3}} e^{A(\pi r_c)} \exp{[\frac{l^2}{3}\exp{(-2u\pi r_c)}]}\nonumber\\
   &[&1 + \exp{(\frac{2l^2}{3}e^{-u\pi r_c} - \frac{2m_5}{k})}]\nonumber\\
   &\bigg(&\frac{e^{(k+2m_5)\pi r_c}}{e^{(k+2m_5)\pi r_c} -1}\bigg)
   \label{coupling 4}
  \end{eqnarray}
  for left handed chiral mode and,
  \begin{eqnarray}
  \lambda_R&=&\sqrt{\frac{k}{24M^3}} e^{A(\pi r_c)} \exp{[\frac{l^2}{3}\exp{(-2u\pi r_c)}]}\nonumber\\
   &[&1 + \exp{(\frac{2l^2}{3}e^{-u\pi r_c} + \frac{2m_5}{k})}]\nonumber\\
   &\bigg(&\frac{e^{(k-2m_5)\pi r_c}}{e^{(k-2m_5)\pi r_c} -1}\bigg)
   \label{coupling 5}
  \end{eqnarray}
  for right handed mode.\\
It is evident that the effective radion-fermion coupling increases (for both left 
and right chiral mode) with the backreaction 
parameter. It is expected because the peak of both left and right chiral wave function 
get shifted towards the visible brane as the backreaction parameter increases.





\subsection{Massive KK mode}
In this section, we study the localization of higher Kaluza-Klein modes. For massive KK 
modes, equation of motion for fermionic wave function is given as,
\begin{eqnarray}
   &\exp&{(-ky-\frac{l^2}{6}e^{-2uy})} [\pm(\partial_{y} - 2k + \frac{2l^2u}{3}e^{-2uy})\nonumber\\ 
   &+&m_5]\xi^n_{R,L}(y) = -m_{n}\xi^n_{L,R}(y)
   \label{eom for wave function2}
  \end{eqnarray}
  
  Recall that $m_n$ is the mass of nth KK mode. Using the rescaling $\tilde{\xi}_{L,R} = e^{\frac{5}{2}A(y)}\xi_{L,R}$, 
  we find that the two helicity states, $\xi_L$ and $\xi_R$ satisfy the same equation of motion and is given by,
  \begin{eqnarray}
   &\tilde{\xi}^n&''(y) + [-\frac{k^2}{4}(1 + \frac{l^2}{3}\frac{u}{k}) + (m_n^2 - \frac{k^2}{4}(1 + \frac{l^2}{3}\frac{u}{k})\nonumber\\
   &-&m_5^2) \exp{(2ky+\frac{l^2}{3}e^{-2uy})}] \tilde{\xi}^n_{L,R}(y) = 0
   \label{reduced eom}
  \end{eqnarray}
  
  Solution of eqn.(\ref{reduced eom}) is given by Bessel function as follows :
  \begin{eqnarray}
   \xi^n_{L,R}(y)&=&\sqrt{k}\exp{(- \frac{5l^2}{12}e^{-2uy})}\nonumber\\
   &\Gamma&(1+\frac{\sqrt{3k+l^2u}}{4\sqrt{3k}}) BesselI[-\frac{\sqrt{3k+l^2u}}{4\sqrt{3k}} ,\nonumber\\ 
   &e&^{-ky}\frac{\sqrt{3k^2+12m_n^2+12m_5^2+kl^2u}}{4\sqrt{3}k}]
   \label{solution3}
  \end{eqnarray}
  
  The mass spectrum can be obtained from the requirement that the wave function is 
  well behaved on the brane. Demanding the continuity of $\xi_{L,R}$ at $y=0$ and at $y=\pi r_c$ 
  gives the mass term as follows :
   \begin{equation}
    m_n^2 = e^{-2A(\pi)} [k^2(n^2 + 2n + 1) + m_5^2]
    \label{mass spectrum}
   \end{equation}
   where $n=1,2,3....$. Now from the requirement of solving the gauge hierarchy 
   problem, the warp factor at TeV brane acquires the value as $A(\pi)=36$ 
   which produces a large suppression in the right hand side of eqn. (\ref{mass spectrum}) 
   through the exponential factor. Since $k, m_5\sim M$, the mass of KK modes ($n=1,2,3..$) comes at TeV scale.\\
   Using the solution of $\xi^n_{L,R}(y)$ (in eqn. (\ref{solution3})), we determine the coupling between 
   massive KK fermion modes and the radion field, given by
   \begin{eqnarray}
    \lambda^{(n)}&=&\sqrt{\frac{k}{24M^3}} e^{A(\pi)} \exp{(- \frac{5l^2}{6}e^{-2u\pi r_c})}\nonumber\\
    &\bigg(&\Gamma(1+\frac{\sqrt{3k+l^2u}}{4\sqrt{3k}}) BesselI[-\frac{\sqrt{3k+l^2u}}{4\sqrt{3k}} ,\nonumber\\ 
   &e&^{-k\pi r_c}\frac{\sqrt{3k^2+12m_n^2+12m_5^2+kl^2u}}{4\sqrt{3}k}]\bigg)^2
   \label{coupling}
   \end{eqnarray}
   where $\lambda^{(n)}$ is the coupling between $n$ th KK fermion mode and the radion field. 
   Eqn. (\ref{coupling}) clearly indicates that $\lambda^{(n)}$ decreases with increasing backreaction parameter.

\begin{figure}[!h]
\begin{center}
 \centering
 \includegraphics[width=3.0in,height=2.0in]{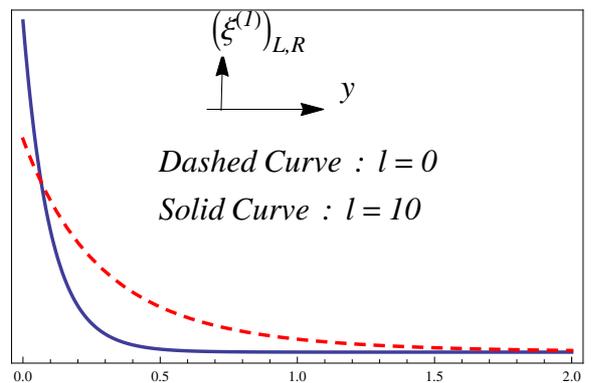}
 \caption{$\xi^1_{L,R}$ vs $y$ for $k=1$, $\frac{u}{k}=0.2$ and $\frac{m_5}{k}=0.5$}
 \label{plot massive mode}
\end{center}
\end{figure}

   Eqn.(\ref{solution3}) indicates the relation between $\xi_{L,R}$ and $y$ for various values of $l$ 
   from which one can find the dependence of localization 
   for massive KK fermion modes on the backreaction parameter. 
   From this, the behaviour of the first KK mode ($n=1$) is described in figure (\ref{plot massive mode}).
   
Figure (\ref{plot massive mode}) clearly depicts that the wave function for 
first massive KK mode gets more and more localized near Planck brane with increasing 
value of backreaction parameter. As a result, the coupling parameter decreases near the visible 
brane as the backreaction parameter increases.\\  
Moreover, it can also be shown (from eqn.(\ref{solution3})) that as the order of KK mode increases from $n=1$, 
the localization of fermions becomes sharper near Planck brane.\\

Before concluding, it may be mentioned that the bulk fermion mass term ($m_5$) also affects the localization of fermion field. 
Using the solution of $\xi_{L,R}(y)$ presented in eqn. (\ref{solution2}), it can be shown that for a fixed value of backreaction 
poaremeter, the left chiral mode of zeroth KK fermion has higher peak values on TeV brane as 
the bulk fermion mass increases where as the right chiral mode 
shows the reverse nature, which is in agreement with \cite{jm}.

\section{Conclusion}
We consider a five dimensional AdS compactified warped geometry model with two 3-branes 
embedded within the spacetime. For the purpose of modulus stabilization, 
a massive scalar field is invoked in the bulk and its backreaction on spacetime 
geometry is taken into account. In this scenario, we study how the backreaction parameter affects 
the localization of a bulk fermion field within the entire spacetime. Moreover, we 
also explore how the fermion localization depends on the bulk mass parameter. Our 
findings are as follows :

\begin{enumerate}
 \item For massless KK mode - 
 \begin{itemize}
  \item In the absence of bulk fermion mass, left and right chiral modes can be localized at different 
  regions in the spacetime by adjusting the value of backreaction parameter ($l$). However, the localization 
  of both the chiral modes becomes sharper near TeV brane as the value of $l$ increases.
  
  \item In the presence of non-zero bulk fermion mass, the left as well as right mode get more and more 
  localized as the backreaction parameter becomes larger. Correspondingly the overlap of fermion 
  wave function with the visible brane increases with $l$, which is depicted in figure (\ref{plot bulk mass left various l zoom}) 
  and figure (\ref{plot bulk mass right various l zoom}).
  
  \item The effective coupling between radion and zeroth order fermionic KK mode is obtained 
  (in eqn. (\ref{coupling 4}) and eqn. (\ref{coupling 5})). 
  It is found that the radion-fermion coupling (for both left and right chiral mode) increases 
  with the increasing value of backreaction parameter. This is a direct consequence of the fact 
  that the peak of the left and right chiral mode get shifted towards the visible brane 
  as the backreaction parameter increases. This in turn enhances the radion to fermion decay 
  amplitude.
  
  \item For a fixed value of backreaction parameter, the left chiral mode has higher peak values on TeV brane 
  as the bulk fermions become more and more massive where as the right chiral mode shows a reverse nature. 
 \end{itemize}
 
 \item For massive KK mode -
 \begin{itemize}
  \item The requirement of solving the gauge hierarchy problem confines the mass 
  of higher KK modes at TeV scale. Moreover the mass squared gap ($\Delta m_n^2 =m_{n+1}^2-m_n^2$) depends linearly 
  on $n$ which is also evident from the mass spectrum in eqn. (\ref{mass spectrum}).
  
  \item The coupling between radion and massive KK fermionic mode is determined in eqn. (\ref{coupling}). 
  It is found that the coupling parameter decreases with the increasing backreaction parameter.
  
  \item From the perspective of localization scenario, the wave function of massive KK 
  modes are localized near Planck brane which  
  increases with the order of KK mode. As a result, the couplings of the massive KK fermionic modes 
  with the visible brane matter fields become extremely weak and therefore drastically reduces the possibility 
  of finding the signatures of such massive fermion KK modes on TeV brane.
 \end{itemize}

\end{enumerate}

\end{document}